\documentclass[12pt]{article}
\begin{document}
\title{Linear viscoelasticity of polyolefin melts:
the effects of temperature and chain branching}

\author{A.D. Drozdov\footnote{Corresponding author;
Fax: (304) 293 4139;
E--mail: Aleksey.Drozdov@mail.wvu.edu}\hspace*{1 mm},
S. Agarwal and R.K. Gupta\\
Department of Chemical Engineering\\
West Virginia University\\
P.O. Box 6102\\
Morgantown, WV 26506, USA}
\date{}
\maketitle

\begin{abstract}
Observations are reported in isothermal torsional
oscillation tests on melts of isotactic polypropylene
(iPP) and low-density polyethylene (LDPE)
in the intervals of temperature between 190
and 250~$^{\circ}$C (iPP) and between 120 and
190~$^{\circ}$C (LDPE).
With reference to the concept of transient networks,
constitutive equations are developed for the
viscoelastic response of polymer melts at three-dimensional
deformations with small strains.
A melt is treated as an equivalent network of strands
bridged by temporary junctions (entanglements and physical
cross-links whose life-times exceed the characteristic
time of deformation).
The time-dependent behavior of the network is
modelled as detachment of active strands from
their junctions and merging of dangling strands
with the network.
The network is assumed to be strongly heterogeneous
in the sense that different junctions have different
activation energies for separation of strands.
The stress--strain relations involve three adjustable
parameters (the plateau modulus,
the average activation energy for rearrangement
of strands and the standard deviation of activation
energies) that are determined by matching the dependencies
of storage and loss moduli on frequency of oscillations.
The difference in the effects of temperature on the
material constants of iPP and LDPE is associated with
the difference in their molecular architecture.
\end{abstract}

\newpage

\section{Introduction}

This paper is concerned with the effect of temperature
on the viscoelastic response of melts of isotactic polypropylene
(iPP) and low-density polyethylene (LDPE).
The choice of these polyolefins for the investigation
may be explained by two reasons.
First, polyethylene and polypropylene are conventional
polymers widely used in industrial applications:
oriented films for packaging,
reinforcing fibres,
non-woven fabrics,
pipes, etc.,
as well as components of copolymers and blends with
improved mechanical properties.
Secondly, these polymers have a similar structure
of backbones (polypropylene differs from polyethylene by
the presence of methyl side-groups only), but quite different
molecular architecture (relatively short-branched chains
in iPP versus long-branched chains in LDPE).
The aim of the present work is to shed some light on
relations between the micro-structure of an ensemble
of macromolecules on the one hand,
and the time-dependent behavior of polymer melts observed
in conventional torsional oscillatory tests, on the other.
In particular, we focus on the effects of long chain branches
on the evolution of (i) the shear modulus of a melt
and (ii) its characteristic relaxation time with temperature.

According to the theory of entropic elasticity (Ferry, 1980),
the elastic modulus $\mu$ of a flexible chain linearly increases
with the absolute temperature $T$,
\begin{equation}
\mu=\mu_{0} k_{\rm B}T,
\end{equation}
where $k_{\rm B}$ is Boltzmann's constant,
and the dimensionless coefficient $\mu_{0}$ is of order of unity
(this coefficient is determined by an averaging method
in the calculation of the chain's entropy).
As conventional concepts in rubber elasticity disregard
interactions between chains (these interactions are accounted
for by means of the incompressibility condition),
the shear modulus of an ensemble of polymer chains
equals the product of the concentration of strands (segments
of macromolecules between contiguous junctions) per unit
mass $N$ by the elastic modulus of an individual chain $\mu$,
\begin{equation}
G=\rho \mu N,
\end{equation}
where $\rho$ stands for mass density.
For linear chains, the number of strands per unit mass $N$
equals the product of the number of chains per unit mass ${\cal N}$
by the average number of strands per chain $\beta$,
\begin{equation}
N=\beta {\cal N},
\qquad
\beta=\frac{L}{L_{\rm e}}.
\end{equation}
Here $L$ is the average contour length of a chain,
and $L_{\rm e}$ is the average contour length between
entanglements.
Combining Eqs. (1) to (3), we arrive at the conventional
formula
\begin{equation}
G=\rho \mu_{0} k_{\rm B} {\cal N} L\frac{T}{L_{\rm e}}.
\end{equation}
Experimental data demonstrate a pronounced decrease of
the apparent elastic modulus (defined, e.g., as the
storage modulus $G^{\prime}$ measured at the maximal
frequency $\omega_{\max}$ available in an experiment)
with temperature $T$ (see Figures 1 and 3 below).
As all parameters on the right-hand side of Eq. (4)
but the ratio $T/L_{\rm e}$ are temperature independent
(insignificant changes in density with temperature
are disregarded in the present study),
this implies that the average contour length between
entanglements $L_{\rm e}$ strongly increases with
temperature,
\begin{equation}
L_{\rm e}(T)=C l_{\rm e}(T),
\qquad
l_{\rm e}(T)=\frac{T}{G(T)},
\qquad
C= \rho \mu_{0} k_{\rm B} {\cal N} L .
\end{equation}
For an ensemble of linear chains, the growth of $L_{\rm e}$
with temperature $T$ may be attributed to an increase in
the average inter-chain distance driven by activation of
thermal fluctuations.
The same process determines an increase in the specific
volume (length) with temperature,
which is described by the conventional equation
\begin{equation}
\frac{1}{l}\frac{dl}{dT}=\alpha,
\end{equation}
where $l$ denotes a characteristic length,
and $\alpha$ is the coefficient of linear thermal expansion.
Based on this similarity, we propose to describe changes
in $l_{\rm e}(T)$ by the differential equation similar to Eq. (6),
\begin{equation}
\frac{1}{l_{\rm e}-l_{\rm e}^{(0)}}\frac{dl_{\rm e}}{dT}
=\alpha_{\rm e},
\end{equation}
where $\alpha_{\rm e}$ stands for an analog of the
coefficient of linear thermal expansion for the average
length between entanglements,
and $l_{\rm e}^{(0)}$ characterizes the contour length
between entanglements at relatively low temperatures $T$.
Solving Eq. (7) for a constant coefficient $\alpha_{\rm e}$,
we find that
\begin{equation}
l_{\rm e}(T)=l_{\rm e}^{(0)}+l_{\rm e}^{(1)}
\exp \Bigl ( \frac{T}{T_{\rm e}}\Bigr ),
\qquad
T_{\rm e}=\frac{1}{\alpha_{\rm e}}.
\end{equation}
Combining Eqs. (5) and (8), we arrive at the formula for
the shear modulus $G$ as a function of temperature $T$,
\begin{equation}
G(T)=\frac{T}{l_{\rm e}^{(0)}+l_{\rm e}^{(1)}
\exp ( T/T_{\rm e})}.
\end{equation}
It is worth noting that an equation similar to Eq. (9)
(where the coefficient $l_{\rm e}^{(0)}$ is absent)
has recently been proposed by Wood-Adams and Costeux (2001)
based on another physical ground.

A substantial increase in the entanglement distance
$l_{\rm e}$ with temperature $T$ has been observed
by Richter et al. (1990, 1993)
about a decade ago by using neutron spin-echo
spectroscopy (NSES) in polyisoprene, polybutadiene
and a poly(ethylene--propylene) alternative copolymer.
To the best of our knowledge, no attempts have been made,
however, to model the growth of $l_{\rm e}(T)$
with the help of differential equations,
as well as to evaluate their parameters by matching observations
in rheological tests.
The latter may be explained by the use of conventional
methods (Ferry, 1980)
for determining the plateau modulus $G^{\circ}$
(an analog of the instantaneous modulus $G$).
These methods are grounded on shifting the graphs
of the storage modulus $G^{\prime}$ measured at various
temperatures $T$ and plotted in the double-logarithmic
coordinates as functions of frequency $\omega$ along both
axes: no restrictions are imposed on the shift along
the abscissa axis (which characterizes the shift factor),
whereas an appropriate shift along the ordinate axis
is presumed to be proportional to the logarithm of
temperature $T$, which is tantamount to formula (2)
with a constant $N$.

The objective of this study is two-fold:
\begin{itemize}
\item
to develop a constitutive model for the isothermal
time-dependent response of a polymer melt at small
strains that involves only three material constants
(including the instantaneous shear modulus $G$),

\item
to report experimental data in torsional oscillation
tests on isotactic polypropylene and low-density
polyethylene at various temperatures $T$,
to determine adjustable parameters in the stress--strain
relations by fitting the observations,
and to verify Eq. (9) by comparing the experimental data
with the results of numerical simulation.
\end{itemize}
To develop constitutive equations tractable from the
mathematical standpoint, we adopt a homogenization concept.
According to it, a complicated micro-structure of
a polymer melt is replaced by an equivalent phase,
whose response captures essential features of the
mechanical behavior.
Following common practice (Sweeney et al., 1999),
a network of chains is chosen as an equivalent phase.

The time-dependent behavior of melts is modelled within
the concept of transient networks
(Green and Tobolsky, 1946;
Yamamoto, 1956;
Lodge, 1968;
Tanaka and Edwards, 1992).
A polymer melt is thought of as a network of
strands bridged by temporary junctions (entanglements
and physical cross-links whose life-time exceeds
the characteristic time of rheological tests).
It is assumed that active strands (whose ends are linked
to contiguous junctions) separate from these junctions,
while dangling strands (that have a free end) merge with
the network.
Detachment and attachment events occur at random
times as appropriate strands are excited by thermal
fluctuations.
Following Drozdov and Christiansen (2003),
we assume the network to be strongly heterogeneous in
the sense that different junctions have different
activation energies for detachment of active strands.
The inhomogeneity of the network is attributed to
(i) density fluctuations in the ensemble of macromolecules,
and (ii) the fact that the life-time of entanglements
created by long branches as well as formed by backbones but
located in the neighborhoods of chain's ends is smaller
than that of knots between backbones of different macromolecules
entangled in the vicinity of their middle-points.
The distribution of active strands with various activation
energies for rearrangement is described by
the random energy model (Derrida, 1980).

The stress--strain relations involve three adjustable parameters:
(i) the instantaneous modulus $G$,
(ii) the average activation energy for separation of strands
from temporary junctions $V_{\ast}$,
and (iii) the standard deviation of activation energies
$\Sigma_{\ast}$,
which, in general, depend on temperature $T$.
These quantities are found by fitting experimental data
for the storage and loss moduli in shear oscillatory tests at
each temperature separately.
This procedure provides an opportunity to evaluate
the dependence $G(T)$ and to verify Eq. (9).

The exposition is organized as follows.
Experimental data in torsional oscillation tests
are reported in Section 2.
Constitutive equations for an heterogeneous transient
network of strands at three-dimensional deformations
are derived in Section 3.
Adjustable parameters in the stress--strain relations
are determined in Section 4.
A brief discussion of our findings is given in Section 5.
Some concluding remarks are formulated in Section 6.

\section{Experimental procedure}

Isotactic polypropylene PP 1012 (density 0.906 g/cm$^{3}$,
melt flow rate 1.2 g/10 min) was purchased from
BP Amoco Polymers, Inc.
Low-density polyethylene Huntsman PE 1020
(density 0.923 g/cm$^{3}$, melt flow rate 2.0 g/10 min)
was supplied by GE Company.
Granules were dried at the temperature $T=100$~$^{\circ}$C
for 12 h prior to molding.
Circular plates with radius 62 mm and thickness 3 mm
were molded in injection-molding machine
Battenfeld 1000/315 CDC (Battenfeld).
Specimens for rheological tests (with diameter 30 mm)
were cut from the plates.

To evaluate melting temperatures of iPP
and LDPE, DSC (differential scanning calorimetry)
measurements were carried out by using DSC 910S apparatus
(TA Instruments).
The calorimeter was calibrated with indium as a standard.
Two specimens of each polymer with weights of approximately
15 mg were tested with a heating rate of 10 K/min from room
temperature to 200~$^{\circ}$C.
The melting temperatures $T_{\rm m}=172$ (iPP)
and $T_{\rm m}=113$~$^{\circ}$C (LDPE)
were determined as the point corresponding to
the peaks on the melting curves.

Rheological tests were performed by using RMS-800 rheometric
mechanical spectrometer with parallel disks (diameter 25 mm,
gap length 2 mm) at the temperatures
$T=190$, 210, 230 and 250~$^{\circ}$C (iPP)
and $T=120$, 130, 140, 150, 160, 170, 180 and 190~$^{\circ}$C
(LDPE).
Given a temperature $T$, at least two dynamic tests were
carried out on different samples.
The shear storage modulus $G^{\prime}$ and
the shear loss modulus $G^{\prime\prime}$
were measured in oscillation tests (the frequency-sweep
mode) with the amplitude of 15 \% and various frequencies
$\omega$ ranging from 0.1 to 100 rad/s.
Our choice of the amplitude of oscillations was based
by the following requirements:
(i) mechanical tests were performed in the region of
linear viscoelasticity, and (ii) the torque
was less than its ultimate value 0.2 N$\cdot$m.
The limitation on the minimum frequency of oscillations
was imposed by two conditions: (i) the torque exceeded
its minimum value $2.0\cdot 10^{-4}$ N$\cdot$m,
and (ii) the duration of a test did not exceed 20 min,
which ensured that thermal degradation of polymers
at elevated temperatures may be disregarded.
To check that the storage and loss moduli were not affected
by the strain amplitude, several tests were repeated with
the amplitude of 5 \%; no changes in dynamic moduli were
observed.
The temperature in the chamber was controlled with
a standard thermocouple that indicated that the temperature
of specimens remained practically constant (with the
accuracy of $\pm 1.0$~$^{\circ}$C).

Each test was performed on a new sample.
The specimen was thermally equilibrated
in the spectrometer (during 5 min),
the gap length was reduced to 2 mm,
an extraneous material was carefully removed,
and the storage and loss moduli were measured
at various frequencies $\omega$ starting from
the lowest one.

The storage $G^{\prime}$ and loss $G^{\prime\prime}$
moduli are depicted versus the logarithm ($\log=\log_{10})$
of frequency $\omega$ in Figures 1 and 2 for iPP
and in Figures 3 and 4 for LDPE.
Conventional semi-logarithmic plots are used to characterize
changes in these quantities with frequency.
According to Figures 1 to 4, the dependencies of storage
and loss moduli on frequency of oscillations
have similar shapes for both polyolefins.
Given a temperature $T$, the storage modulus $G^{\prime}$
and the loss modulus $G^{\prime\prime}$ strongly increase
with frequency $\omega$.
For a fixed frequency $\omega$, the dynamic moduli
pronouncedly decrease with temperature $T$.

\section{Constitutive equations}

With reference to the concept of transient networks,
a polymer melt is thought of as an equivalent network of
strands bridged by temporary junctions.
A strand whose ends are linked to contiguous junctions is
treated as an active one.
When an end of an active strand separates from a junction,
the strand is transformed into the dangling state.
When a free end of a dangling strand captures a nearby
junction, the strand returns into the active state.
Separation of active strands from their junctions and
merging of dangling strands with the network occur at
random times when the strands are excited by thermal
fluctuations.
According to the theory of thermally-activated processes
(Eyring, 1936), the rate of detachment of strands from
temporary junctions $\Gamma$ is governed by the equation
\begin{equation}
\Gamma=\Gamma_{0} \exp \Bigl (-\frac{\bar{v}}{k_{\rm B}T}\Bigr ),
\end{equation}
where $\Gamma_{0}$ is the attempt rate (the number of
separation events per strand per unit time),
$k_{\rm B}$ is Boltzmann's constant,
$T$ is the absolute temperature,
and $\bar{v}\geq 0$ is the activation energy for separation
of an active strand.
In what follows, we set $\Gamma_{0}=10^{11}$ s$^{-1}$,
which corresponds to the characteristic relaxation rate
at the monomeric scale (deGennes, 1979).

For isothermal deformation at a temperature $T$,
we introduce the dimensionless activation energy
\begin{equation}
v=\frac{\bar{v}}{k_{\rm B}T},
\end{equation}
and present Eq. (10) in the form
\begin{equation}
\Gamma(v)=\Gamma_{0}\exp (-v).
\end{equation}
To describe the time-dependent response of a melt,
we suppose that different junctions are characterized by
different dimensionless activation energies $v$
(Drozdov and Christiansen, 2003).
The distribution of active strands in a transient network
is determined by the number of active strands per unit mass
$N$ and the distribution function $p(v)$.
The quantity $Np(v)dv$ equals the number
of active strands per unit mass linked to junctions with
the dimensionless activation energies $u$ belonging
to the interval $[v,v+dv]$.

Separation of active strands from temporary junctions
and merging of dangling strands with the network are entirely
described by the function $n(t,\tau,v)$
that equals the number (per unit mass) of active strands
at time $t\geq 0$ linked to temporary junctions with
activation energy $v$ which have last merged with the network
before instant $\tau\in [0,t]$.

The quantity $n(t,t,v)$ equals the number of active strands
(per unit mass) with the activation energy $v$ at time $t$,
\begin{equation}
n(t,t,v)=Np(v).
\end{equation}
The function
\begin{equation}
\gamma(\tau,v)=\frac{\partial n}{\partial \tau}(t,\tau,v) \biggl
|_{t=\tau}
\end{equation}
determines the rate of reformation for dangling chains:
the amount $\gamma(\tau,v)d\tau$ equals the number of dangling
strands (per unit mass) that merge with temporary junctions
with activation energy $v$ within the interval $[\tau,\tau+d\tau]$.
The quantity
\[
\frac{\partial n}{\partial \tau} (t,\tau,v)\;d\tau
\]
is the number of these strands that have not separated from
their junctions during the interval $[\tau, t]$.
The amount
\[
-\frac{\partial n}{\partial t} (t,0,v)\;dt
\]
is the number of active strands (per unit mass) that detach (for
the first time) from the network within the interval $[t,t+dt ]$,
while the quantity
\[
-\frac{\partial^{2} n}{\partial t\partial \tau} (t,\tau,v)\;dtd\tau
\]
equals the number of strands (per unit mass) that have last merged
with the network within the interval $[\tau,\tau+d\tau ]$ and
separate from the network (for the first time after merging)
during the interval $[t,t+dt ]$.

The rate of detachment $\Gamma$ is defined as the ratio of
the number of active strands that separate from temporary
junctions per unit time to the total number of active strands.
Applying this definition to active strands that were connected
with the network at the initial instant $t=0$,
and to those that merged with the network within the interval
$[\tau,\tau +d\tau ]$, we arrive at the differential equations
\begin{equation}
\frac{\partial n}{\partial t}(t,0,v)
= - \Gamma(v) n(t,0,v),
\qquad
\frac{\partial^{2} n}{\partial t\partial \tau}(t,\tau,v)
= - \Gamma(v) \frac{\partial n}{\partial \tau}(t,\tau,v).
\end{equation}
Integration of Eq. (15) with initial conditions (13) (where we set
$t=0$) and (14) implies that
\begin{equation}
n(t,0,v) = N p(v) \exp \Bigl [-\Gamma(v)t \Bigr ],
\qquad
\frac{\partial n}{\partial \tau}(t,\tau,v)
= \gamma(\tau,v) \exp \Bigl [-\Gamma(v)(t-\tau) \Bigr ].
\end{equation}
To exclude the function $\gamma(t,v)$ from Eq. (16), we use the
identity
\begin{equation}
n(t,t,v)=n(t,0,v)+\int_{0}^{t}
\frac{\partial n}{\partial \tau}(t,\tau,v) d\tau.
\end{equation}
Substitution of expressions (13) and (16) into Eq. (17) results in
\begin{equation}
N p(v) = N p(v)
\exp \Bigl [-\Gamma(v)t \Bigr ]
+\int_{0}^{t} \gamma(\tau,v)\exp \Bigl [-\Gamma(v)(t-\tau)\Bigr ]
d\tau.
\end{equation}
The solution of linear integral equation (18) reads
$\gamma(t,v)=N p(v)\Gamma(v)$.
It follows from this equality and Eq. (16) that
\begin{equation}
\frac{\partial n}{\partial \tau}(t,\tau,v)
= N p(v) \Gamma(v) \exp \Bigl [-\Gamma(v)(t-\tau) \Bigr ].
\end{equation}

We adopt the conventional assumptions that
(i) the excluded-volume effect and other multi-chain effects
are screened for individual strands by surrounding
macromolecules,
(ii) the energy of interaction between strands can be taken
into account with the help of the incompressibility
condition,
and (iii) thermal oscillations of junctions can be disregarded,
and the strain tensor for the motion of junctions at
the micro-level coincides with the strain tensor for
macro-deformation.

At isothermal deformation with small strains, a strand
is treated as an isotropic incompressible medium
with the strain energy
\[
w_{0}=\mu \hat{e}^{\prime}:\hat{e}^{\prime},
\]
where $\hat{e}$ is the strain tensor for transition from
the reference (stress-free) state of the strand to its
deformed state,
the average elastic modulus $\mu$ is given by Eq. (1),
the prime stands for the deviatoric component of a tensor,
and the colon denotes convolution of two tensors.

According to the affinity hypothesis, the strain energy
$\bar{w}_{0}(t,0)$ of an active strand that has not separated
from the network during the interval $[0,t]$ reads
\[
w(t,0)=\mu \hat{\epsilon}^{\prime}(t):\hat{\epsilon}^{\prime}(t),
\]
where $\hat{\epsilon}(t)$ is the strain tensor for transition from
the initial (stress-free) state of the network to its deformed
state at time $t$.
With reference to Tanaka and Edwards (1992), we suppose that
stress in a dangling strand totally relaxes before this strand
captures a new junction.
This implies that the stress-free state of an active strand
that merges with the network at time $\tau\geq 0$ coincides
with the deformed state of the network at that instant.
The strain energy of an active strand that has
last merged with the network at time $\tau\in [0,t]$ is given by
\[
w(t,\tau)=\mu \Bigl [ \hat{\epsilon}(t)
-\hat{\epsilon}(\tau)\Bigl ]^{\prime}:
\Bigl [ \hat{\epsilon}(t)-\hat{\epsilon}(\tau)\Bigl ]^{\prime}.
\]
Multiplying the strain energy per strand by the number of
active strands per unit mass and summing the mechanical
energies of active strands linked to temporary junctions with
various activation energies,
we find the strain energy per unit mass of an equivalent network
\begin{equation}
W(t)=\mu \int_{0}^{\infty} \biggl \{ n(t,0,v)
\hat{\epsilon}^{\prime}(t):\hat{\epsilon}^{\prime}(t)
+\int_{0}^{t} \frac{\partial n}{\partial \tau}(t,\tau,v)
\Bigl [ \hat{\epsilon}(t)-\hat{\epsilon}(\tau)\Bigl ]^{\prime}
:\Bigl [ \hat{\epsilon}(t)-\hat{\epsilon}(\tau)\Bigl ]^{\prime}
d\tau \biggr \}dv.
\end{equation}
Differentiating Eq. (20) with respect to time $t$
and using Eqs. (16) and (19), we arrive at the formula
\begin{equation}
\frac{dW}{dt}(t)=
\hat{A}^{\prime}(t):\frac{d\hat{\epsilon}^{\prime}}{dt}(t)-B(t),
\end{equation}
where
\begin{eqnarray}
\hat{A}(t) &=& 2\mu N \biggl \{ \hat{\epsilon}(t)
-\int_{0}^{t} \hat{\epsilon}(\tau) d\tau
\int_{0}^{\infty} \Gamma(v)\exp \Bigl [-\Gamma(v)(t-\tau)\Bigr ]
p(v) dv \biggr \},
\\
B(t) &=& \mu \int_{0}^{\infty} \Gamma(v) \biggl \{ n(t,0,v)
\hat{\epsilon}^{\prime}(t):\hat{\epsilon}^{\prime}(t)
\nonumber\\
&& +\int_{0}^{t} \frac{\partial n}{\partial \tau}(t,\tau,v)
\Bigl [ \hat{\epsilon}(t)-\hat{\epsilon}(\tau)\Bigl ]^{\prime}
:\Bigl [ \hat{\epsilon}(t)-\hat{\epsilon}(\tau)\Bigl ]^{\prime}
d\tau \biggr \} dv \geq 0.
\end{eqnarray}
For isothermal deformation of an incompressible medium,
the Clausius--Duhem inequality reads
\[
Q=-\frac{dW}{dt}+\frac{\hat{\sigma}^{\prime}}{\rho}:
\frac{d\hat{\epsilon}^{\prime}}{dt} \geq 0,
\]
where $Q$ is internal dissipation per unit mass,
and $\hat{\sigma}$ stands for the stress tensor.
Substitution of Eq. (21) into this equation implies that
\begin{equation}
Q(t)=\frac{1}{\rho} \Bigl [\hat{\sigma}^{\prime}(t)
-\rho \hat{A}^{\prime}(t)\Bigr ]: \frac{d\hat{\epsilon}^{\prime}}{dt}(t)
+B(t)\geq 0.
\end{equation}
As the function $B(t)$ is non-negative, see Eq. (23),
dissipation inequality (24) is satisfied, provided that the
expression in the square brackets vanishes.
This assertion together with Eq. (22) results
in the constitutive equation
\begin{equation}
\hat{\sigma}(t) = -P(t)\hat{I}+2G \biggl \{ \hat{\epsilon}^{\prime}(t)
-\int_{0}^{t} \hat{\epsilon}^{\prime}(\tau) d\tau
\int_{0}^{\infty} \Gamma(v)\exp \Bigl [-\Gamma(v)(t-\tau)\Bigr ]
p(v) dv \biggr \},
\end{equation}
where $P(t)$ is pressure,
$\hat{I}$ is the unit tensor,
and the shear modulus $G$ is determined by Eq. (2).

Formula (25) describes the time-dependent response of an
equivalent network at arbitrary three-dimensional deformations
with small strains.
This equation implies that in a shear test with
$\hat{\epsilon}(t)=\epsilon(t){\bf e}_{1}{\bf e}_{2}$,
where $\epsilon(t)$ is the shear strain,
and ${\bf e}_{m}$ ($m=1,2,3$) are unit vectors of a Cartesian
frame,
the shear stress $\sigma(t)$ reads
\begin{equation}
\sigma(t) = 2G \biggl \{ \epsilon(t)
-\int_{0}^{t} \epsilon(\tau) d\tau
\int_{0}^{\infty} \Gamma(v)\exp \Bigl [-\Gamma(v)(t-\tau)\Bigr ]
p(v) dv \biggr \}.
\end{equation}
It follows from Eq. (26) that in a shear oscillation test with
$\epsilon(t)=\epsilon_{0}\exp (i\omega t)$,
where $\epsilon_{0}$ and $\omega$ are the amplitude and frequency of
oscillations, and $i=\sqrt{-1}$, the transient complex modulus
$\bar{G}^{\ast}(t,\omega)=\sigma(t)/(2\epsilon(t))$
is determined by the formula
\[
\bar{G}^{\ast}(t,\omega)=G\biggl \{ 1-\int_{0}^{\infty} \Gamma(v)
p(v) dv \int_{0}^{t}
\exp \Bigl [-\Bigl (\Gamma(v)+i\omega \Bigr )s \Bigr ]ds \biggr \},
\]
where $s=t-\tau$.
This equality implies that the steady-state complex modulus
$G^{\ast}(\omega)=\lim_{t\to\infty} \bar{G}^{\ast}(t,\omega)$
is given by
\[
G^{\ast}(\omega)=G\int_{0}^{\infty}
\frac{i\omega}{\Gamma(v)+i\omega} p(v) dv.
\]
This equality together with Eq. (12) implies that
the steady-state storage $G^{\prime}(\omega)$ and loss
$G^{\prime\prime}(\omega)$ shear moduli read
\begin{eqnarray}
G^{\prime}(\omega) &=& G \int_{0}^{\infty}
\frac{\omega^{2}}{\Gamma_{0}^{2}\exp(-2v)+\omega^{2}}p(v) dv,
\nonumber\\
G^{\prime\prime}(\omega) &=& G \int_{0}^{\infty} \frac{\Gamma_{0}
\exp(-v)\omega}{\Gamma_{0}^{2}\exp(-2v)+\omega^{2}}p(v) dv.
\end{eqnarray}
To fit the experimental data, we adopt the random energy
model (Derrida, 1980) with the quasi-Gaussian distribution
function $p(v)$,
\begin{equation}
p(v) = p_{0}\exp \biggl [
-\frac{(v-V)^{2}}{2\Sigma^{2}}\biggr ]
\quad (v\geq 0),
\qquad
p(v)=0
\quad
(v<0),
\end{equation}
where $V$ and $\Sigma$ are adjustable parameters
(an apparent average activation energy and
an apparent standard deviation of activation
energies, respectively),
and the constant $p_{0}$ is found from the normalization
condition
\begin{equation}
\int_{0}^{\infty} p(v) d v =1.
\end{equation}
The average activation energy for separation of strands
$V_{\ast}$ and the standard deviation of activation
energies $\Sigma_{\ast}$ are determined by the
conventional formulas
\begin{equation}
V_{\ast}=\int_{0}^{\infty} v p(v) dv,
\qquad
\Sigma_{\ast}=\biggl [ \int_{0}^{\infty} (v-V_{\ast})^{2} p(v) dv
\biggr ]^{\frac{1}{2}}.
\end{equation}
Governing equations (27) and (28) involve three material constants:
(i) the instantaneous shear modulus $G$,
(ii) the average activation energy for rearrangement of strands
in a network $V_{\ast}$,
and (iii) the standard deviation of activation energies
$\Sigma_{\ast}$.
For shear oscillatory tests performed at various temperatures
$T$, these quantities become functions of $T$.
Our purpose now is to find these parameters by
matching the observations for $G^{\prime}(\omega)$ and
$G^{\prime\prime}(\omega)$ depicted in Figures 1 to 4.

\section{Fitting of observations}

Each set of the experimental data is fitted separately.
We fix some intervals $[0,V_{\max}]$ and $[0,\Sigma_{\max}]$,
where the ``best-fit" parameters $V$ and $\Sigma$
are assumed to be located,
and divide these intervals into $J$ subintervals by
the points $V^{(i)}=i\Delta V$ and
$\Sigma^{(j)}=j\Delta \Sigma$ ($i,j=1,\ldots,J-1$) with
$\Delta V=V_{\max}/J$ and $\Delta \Sigma=\Sigma_{\max}/J$.
For any pair $\{ V^{(i)}, \Sigma^{(j)} \}$,
the coefficient $p_{0}$ in Eq. (28) is determined from Eq. (29),
where the integral is evaluated numerically by Simpson's
method with 400 points and the step $\Delta v=0.1$.
The integrals in Eq. (27) are calculated by using
the same technique.
The shear modulus $G$ is found by the least-squares method
from the condition of minimum of the function
\[
F=\sum_{\omega_{m}} \biggl \{ \Bigl [
G^{\prime}_{\rm exp}(\omega_{m})
-G^{\prime}_{\rm num}(\omega_{m}) \Bigr ]^{2}
+\Bigl [ G^{\prime\prime}_{\rm exp}(\omega_{m})
-G^{\prime\prime}_{\rm num}(\omega_{m}) \Bigr ]^{2} \biggr \},
\]
where the sum is calculated over all frequencies
$\omega_{m}$ at which the data were collected,
$G_{\rm exp}^{\prime}$ and $G_{\rm exp}^{\prime\prime}$
are the storage and loss moduli measured in a test,
and $G_{\rm num}^{\prime}$ and $G_{\rm num}^{\prime\prime}$
are given by Eq. (27).
The ``best-fit" parameters $V$ and $\Sigma$ are determined
from the condition of minimum of the function $F$
on the set $ \{ V^{(i)}, \Sigma^{(j)} \}$.
After finding the ``best-fit" values $V^{(i)}$ and
$\Sigma^{(j)}$, this procedure is repeated twice
for the new intervals $[ V^{(i-1)}, V^{(i+1)}]$ and
$[ \Sigma^{(j-1)}, \Sigma^{(j+1)}]$,
to ensure an acceptable accuracy of fitting.
Figures 1 to 4 demonstrate good agreement between
the experimental data and the results of numerical
simulation.

After finding the instantaneous shear modulus $G$ of
iPP and LDPE, we calculate the parameter $l_{\rm e}$
by using Eq. (5) and plot this quantity versus temperature
$T$ in Figure 5.
To demonstrate the level of discrepancies between
the values of $l_{\rm e}$ measured on different
specimens, experimental data are presented for two
sets of observations on iPP.
The dependence $l_{\rm e}(T)$ is approximated by Eq. (8),
where the coefficients $l_{\rm e}^{(m)}$ ($m=0,1$)
are determined by the least-squares technique.
Figure 5 shows that Eq. (8) provides fair
approximation of the observations.
The average contour length between entanglements
$l_{\rm e}$ grows with temperature $T$.
The rate of increase in $l_{\rm e}(T)$ is more
pronounced for iPP than for LDPE.

Given adjustable parameters $V$ and $\Sigma$
determined at each test temperature $T$,
we calculate the dimensionless average activation
energy for separation of strands from temporary
junctions $V_{\ast}$ and the dimensionless standard
deviation of activation energies $\Sigma_{\ast}$
by formulas (30), where the integrals are evaluated
numerically.
The average activation energy $\bar{V}_{\ast}$
and the standard deviation of activation energies
$\bar{\Sigma}_{\ast}$ are found from the equations
similar to Eq. (11),
\[
\bar{V}_{\ast}=k_{\rm B}TV_{\ast},
\qquad
\bar{\Sigma}_{\ast}=k_{\rm B}T \Sigma_{\ast}.
\]
The average activation energy $\bar{V}_{\ast}$
is depicted versus temperature $T$ in Figure 6.
The experimental data are approximated by the
linear equation
\begin{equation}
\bar{V}_{\ast}=(V_{0}+V_{1} T)\cdot 10^{-19},
\end{equation}
where the coefficients $V_{m}$ ($m=0,1$) are determined
by the least-squares method.
Figure 6 demonstrates that Eq. (31) correctly describes
the evolution of $\bar{V}_{\ast}$ with temperature.
The average activation energy for rearrangement of strands
$\bar{V}_{\ast}$ grows with temperature for both polymers.
The rate of increase is more pronounced for LDPE than for iPP.
Given a temperature $T$, the average activation energy for
separation of strands in iPP noticeably exceeds that
in LDPE.

The standard deviation of activation energies
$\bar{\Sigma}_{\ast}$ is plotted versus temperature $T$
in Figure 7.
The experimental data are approximated by the
linear function
\begin{equation}
\bar{\Sigma}_{\ast}=(\Sigma_{0}+\Sigma_{1}T)\cdot 10^{-19},
\end{equation}
where the coefficients $\Sigma_{m}$ ($m=0,1$) are determined
by the least-squares technique.
Figure 7 reveals that Eq. (32) adequately describes
changes in $\bar{\Sigma}_{\ast}$ with temperature.
The standard deviation of activation energies
$\bar{\Sigma}_{\ast}$ slightly increases with $T$
for iPP and weakly decreases with temperature for LDPE.
Given a temperature $T$, the standard deviation of
activation energies for LDPE substantially exceeds
that for iPP.

To assess the effect of temperature $T$ on the average
relaxation time $\tau_{0}$, we calculate the modulus of
complex viscosity $\eta$ by the standard formula
\begin{equation}
\eta(\omega)=\biggl [ \Bigl ( G^{\prime}(\omega)\Bigr )^{2}
+\Bigl (\frac{G^{\prime\prime}(\omega)}{\omega}\Bigr )^{2}
\biggr ]^{\frac{1}{2}},
\end{equation}
and find the zero-frequency complex viscosity
$\eta_{0}=\lim_{\omega\to 0}\eta(\omega)$.
It follows from Eqs. (27) and (33) that
\begin{equation}
\eta_{0}=\frac{G}{\Gamma_{0}} \int_{0}^{\infty}
\exp(v) p(v) dv.
\end{equation}
The average relaxation time $\tau_{0}$ is given by
\[
\tau_{0}=\frac{\eta_{0}}{G}.
\]
Substitution of expression (35) into this equality
results in
\begin{equation}
\tau_{0}=\frac{1}{\Gamma_{0}} \int_{0}^{\infty}
\exp(v) p(v) dv.
\end{equation}
For any temperature under consideration $T$, we calculate
the integral in Eq. (35) by Simpson's method with 400 points
and the step $\Delta v=0.1$.
The distribution function $p(v)$ is given by Eq. (28) with
the parameters $V$ and $\Sigma$ found by fitting the
experimental data for $G^{\prime}(\omega)$ and
$G^{\prime\prime}(\omega)$.
The average relaxation time $\tau_{0}$ is plotted
versus temperature $T$ in Figure 8.
The experimental data are approximated by the Arrhenius
dependence
\begin{equation}
\tau_{0}=\tau_{\ast}\exp \Bigl (\frac{E}{RT}\Bigr ),
\end{equation}
where $\tau_{\ast}$ is the average relaxation time at
elevated temperatures ($T\to\infty$),
$E$ is the activation energy,
and $R$ is the universal gas constant.
To match the observations, we present Eq. (36) in the form
\begin{equation}
\ln \tau_{0}=\tau_{0}^{(0)}+\frac{\tau_{0}^{(1)}}{T}
\end{equation}
with
\begin{equation}
\tau_{0}^{(0)}=\ln \tau_{\ast},
\qquad
\tau_{0}^{(1)}=\frac{E}{R}.
\end{equation}
Figure 8 demonstrates that Eq. (37), where the coefficients
$\tau_{0}^{(m)}$ $(m=0,1)$ are determined by the least-squares
technique, adequately describes the experimental data
for both polymers.

\section{Discussion}

Figures 5 shows the average contour length between
entanglements $l_{\rm e}$ strongly increases with
temperature for both polyolefins under investigation.
The characteristic temperature for disentanglement
$T_{\rm e}$ of LDPE exceeds that of iPP by about twice.
To calculate the coefficient $\alpha_{\rm e}$ in Eq. (7),
we use Eq. (8) and the results of numerical analysis
and find $\alpha_{\rm e}=0.023$ K$^{-1}$ for iPP and
$\alpha_{\rm e}=0.013$ K$^{-1}$ for LDPE.
This means that $l_{\rm e}$ changes rather weakly
with temperature for a highly branched polymer melt,
and it is strongly affected by temperature for
a melt with relatively short branches.

This finding may be explained based on the following
scenario.
Slightly above its melting temperature $T_{\rm m}$,
polypropylene chains are closely packed, which implies
that the number of macromolecules that intersect a
volume occupied by an individual chain is relatively
large (a high average number of entanglements per chain
and a small distance between entanglements $l_{\rm e}$).
With the growth of temperature, thermal oscillations
of segments induce loosening of this packaging and an
increase in the occupied volume.
Some macromolecules that crossed the occupied volume
for an individual chain near the melting point are
forced to leave this volume with an increase in $T$.
This results in a substantial decrease in the number
of entanglements per chain and a strong exponential growth
of $l_{\rm e}$.

On the contrary, due to the presence of long branches,
packaging of chains in LDPE at the temperatures $T$
close to its melting temperature $T_{\rm m}$ is relatively
loose.
This poor packaging is associated with formation of
physical cross-links (knots) between long branches and between
long branches and the backbones of polyethylene chains
(in addition to entanglements between backbones).
The growth of temperature induces partial disentanglement
of junctions formed by long branches, which implies
that the main chains become more closely packed
(the latter is reflected by the fact that
the coefficient of thermal expansion of LDPE is negative).
Although the total number of entanglements between chains
is reduced with temperature, the rate of increase in $l_{\rm e}(T)$
in LDPE is weaker than that in iPP, in agreement with
the experimental data depicted in Figure 5.

Figure 6 demonstrates that the average activation energy
for separation of strands from temporary junctions
$\bar{V}_{\ast}$ slightly grows with temperature for iPP.
This increase may be associated with the fact that
disentanglement of chains driven by the growth of
temperature begins with relatively weak junctions
(that are characterized by small activation energies).
The latter means that the average activation energy of
an ensemble of chains increases due to disappearance
of ``weak" junctions untangled under heating.

The rate of growth of the average activation energy
$\bar{V}_{\ast}$ of LDPE with temperature $T$ is
substantially stronger than that of iPP.
This may be explained by the fact that an increase in
temperature induces breakage of physical cross-links between
long branches and between long branches and backbones
in LDPE (which are characterized by relatively low
activation energies).
As a result of thermally-induced destruction of ``weak"
junctions, only relatively strong knots between
backbones survive.
Because these ``strong" entanglements have high activation
energies, the average activation energy of an ensemble
pronouncedly increases with temperature.

According to Figure 6, the average activation energy
$\bar{V}_{\ast}$ of iPP substantially exceeds that of LDPE.
This conclusion appears to be natural, because the
activation energy of a junction is determined by
local properties of entangled chains.
The strength of a knot formed by two chains is relatively low
for LDPE (interactions between backbones of linear chains),
and it is noticeably higher for iPP due
to additional interactions between methyl side-groups
belonging to different chains.

Figure 7 shows that the standard deviations of activation
energies $\bar{\Sigma}_{\ast}$ of iPP and LDPE are weakly
affected by temperature.
The parameter $\bar{\Sigma}_{\ast}$ slightly increases
with temperature for iPP (due to partial disentanglement
of backbones), and it slightly decreases with $T$ for LDPE
(driven by breakage of weak junctions between long branches).
Given a temperature $T$, the standard deviation of
activation energies of LDPE substantially exceeds
$\bar{\Sigma}_{\ast}$ of iPP.
This finding seems quite natural, because the presence of
``weak" physical cross-links between long branches
implies a broad distribution of activation energies for
rearrangement of strands.

Figure 8 reveals that the average relaxation time $\tau_{0}$
grows with temperature for both polymers under consideration.
Given $\tau_{0}^{(1)}$, we calculate the activation energy $E$
from Eq. (38) and find $E=10.4$ kcal/mol for iPP
and $E=18.1$ kcal/mol for LDPE.
These values are in good accord with the activation energies
provided by other researchers for polypropylene:
$E=9.3$ kcal/mol (Eckstein et al., 1998),
$E=9.7-10.0$ kcal/mol (Pearson et al., 1988),
$E=10.0$ kcal/mol (Fujiyama et al., 2002),
and low-density polyethylene:
$E=9.1-13.2$ kcal/mol (Wood-Adams and Costeaux, 2001),
$E=13.5$ kcal/mol (Qiu and Ediger, 2000).

\section{Concluding remarks}

Two series of torsional oscillation tests
have been performed on melts of isotactic
polypropylene and low-density polyethylene
in the range of temperatures between 170
and 250~$^{\circ}$C (iPP)
and between 120 and 190~$^{\circ}$C (LDPE).

With reference to the concept of transient networks,
constitutive equations have been developed for the
viscoelastic response of polymer melts at isothermal
three-dimensional deformations with small strains.
The melt is treated as an equivalent transient network
of strands bridged by temporary junctions.
Its time-dependent behavior is modelled as thermally-activated
separation of active strands from their junctions and
attachment of dangling strands to the network.
Stress--strain relations for an equivalent heterogeneous
network of strands (where different junctions have different
activation energies for rearrangement of strands)
have been derived by using the laws of thermodynamics.
These equations involve three material parameters
that are determined by matching the experimental data
for the storage and loss moduli as functions of frequency
of oscillations.
Fair agreement is demonstrated between the observations
and the results of numerical simulation.

The following conclusions are drawn:
\begin{enumerate}
\item
The average contour length between entanglements
$l_{\rm e}$ grows with temperature $T$ for both
polyolefins under consideration.
Thermally-induced changes in $l_{\rm e}(T)$
are correctly described by Eq. (7).
The relative rate of increase in $l_{\rm e}(T)$ is
higher for iPP than for LDPE.

\item
The average activation energy for rearrangement of
strands $\bar{V}_{\ast}$ grows with temperature
due to breakage of ``weak" junctions between chains.
The rate of increase in $\bar{V}_{\ast}$ for LDPE exceeds
than for iPP (due to thermally-induced destruction
of knots between long branches).
Given a temperature $T$, the average activation energy
of iPP is substantially higher than that of LDPE
(due to local interactions between methyl side-groups).

\item
The standard deviation of activation energies for
separation of active strands $\bar{\Sigma}_{\ast}$
is weakly affected by temperature.
The value of $\bar{\Sigma}_{\ast}$ for LDPE
noticeably exceeds that for iPP (which reflects
the fact that the presence of ``weak" physical
cross-links between long branches in polyethylene
results in a very broad distribution of activation
energies).
\end{enumerate}
An explicit expression (35) is derived for the average
relaxation time $\tau_{0}$.
It is demonstrated that the activation energies of iPP
and LDPE calculated by using Eqs. (35) and (36) are close
to those reported in the literature.

\subsection*{Acknowledgement}

This work was partially supported by the West Virginia
Research Challenge Grant Program.

\newpage
\section*{References}

\begin{description}

\small{

\item
de Gennes PG (1979)
Scaling concepts in polymer physics.
Cornell Univ. Press, Ithaka

\item
Derrida B (1980)
Random-energy model: limit of a family of disordered models.
Phys Rev Lett 45: 79--92

\item
Drozdov AD, Christiansen JdeC. (2003)
The effect of annealing on the viscoplastic response
of semicrystalline polymers at finite strains.
Int J Solids Struct 40: 1337--1367.

\item
Eckstein A, Suhm J, Friedrich C, Maier R-D, Sassmannshausen J,
Bochmann M, M\"{u}lhaupt R (1998)
Determination of plateau moduli and entanglement molecular
weights of isotactic, syndiotactic, and atactic polypropylenes
synthesized with metallocene catalysis.
Macromolecules 31: 1335--1340

\item
Eyring H (1936)
Viscosity, plasticity, and diffusion as examples of absolute
reaction rates.
J Chem Phys 4: 283--291

\item
Ferry JD (1980)
Viscoelastic properties of polymers.
Wiley, New York

\item
Fujiyama M, Kitajima Y, Inata H (2002)
Rheological properties of polypropylenes with different molecular
weight distribution characteristics.
J Appl Polym Sci 84: 2128--2141

\item
Green MS, Tobolsky AV (1946)
A new approach to the theory of relaxing polymeric media.
J Chem Phys 14: 80--92

\item
Lodge AS (1968)
Constitutive equations from molecular network theories
for polymer solutions.
Rheol Acta 7: 379--392

\item
Pearson DS, Fetters LJ, Younghouse LB, Mays JW (1988)
Rheological properties of poly(1,3-di\-meth\-yl-1-butenylene)
and model atactic polypropylene.
Macromolecules 21: 478--484

\item
Qiu X, Ediger MD (2000)
Branching effects on the segmental dynamics of polyethylene
melts.
J Polym Sci Part B: Polym Phys 38: 2634--2643

\item
Richter D, Farago B, Fetters LJ, Huang JS, Ewen B, Lartigue C (1990)
Direct microscopic observation of the entanglement distance
in a polymer melt.
Rhys Rev Lett 64: 1389--1392

\item
Richter D, Farago B, Butera R, Fetters LJ, Huang JS, Ewen B (1993)
On the origins of entanglement constrains.
Macromolecules 26: 795--804

\item
Sweeney J, Collins TLD, Coates PD, Duckett RA (1999)
High temperature large strain viscoelastic behaviour of polypropylene
modeled using an inhomogeneously strained network.
J Appl Polym Sci 72: 563--575

\item
Tanaka F, Edwards SF (1992)
Viscoelastic properties of physically cross-linked networks.
Transient network theory.
Macromolecules 25: 1516--1523

\item
Wood-Adams P, Costeux S (2001)
Thermorheological behavior of polyethylene:
effects of microstructure and long chain branching.
Macromolecules 34: 6281--6290

\item
Yamamoto M (1956)
The visco-elastic properties of network structure.
1. General formalism.
J Phys Soc Japan 11: 413--421

}
\end{description}

\newpage
\section*{List of figures}
\parindent 0 mm
{\small

{\bf Figure 1:} The storage modulus $G^{\prime}$ versus
frequency $\omega$.
Circles: experimental data on iPP at the temperatures
$T=190$, 210, 230 and 250~$^{\circ}$C, from top to bottom,
respectively.
Solid lines: results on numerical simulation
\vspace*{2 mm}

{\bf Figure 2:}
The loss modulus $G^{\prime\prime}$
versus frequency $\omega$.
Circles: experimental data on iPP at the temperatures
$T=190$, 210, 230 and 250~$^{\circ}$C,
from top to bottom, respectively.
Solid lines: results on numerical simulation
\vspace*{2 mm}

{\bf Figure 3:}
The storage modulus $G^{\prime}$
versus frequency $\omega$.
Circles: experimental data on LDPE at the temperatures
$T=120$, 130, 140, 150, 160, 170, 180 and 190~$^{\circ}$C,
from top to bottom, respectively.
Solid lines: results on numerical simulation
\vspace*{2 mm}

{\bf Figure 4:}
The loss modulus $G^{\prime\prime}$
versus frequency $\omega$.
Circles: experimental data on LDPE at the temperatures
$T=120$, 130, 140, 150, 160, 170, 180 and 190~$^{\circ}$C,
from top to bottom, respectively.
Solid lines: results on numerical simulation
\vspace*{2 mm}

{\bf Figure 5:}
The parameter $l_{\rm e}$ versus temperature $T$.
Symbols: treatment of observations on iPP
(unfilled circles) and LDPE (filled circles).
Solid lines: approximation of the experimental data
by Eq. (8).
Curve 1: $l_{\rm e}^{(0)}=1.09$,
$l_{\rm e}^{(1)}=1.04\cdot 10^{-5}$, $T_{0}=43.2$.
Curve 2: $l_{\rm e}^{(0)}=-0.09$,
$l_{\rm e}^{(1)}=1.30\cdot 10^{-3}$, $T_{0}=77.2$
\vspace*{2 mm}

{\bf Figure 6:}
The average activation energy $\bar{V}_{\ast}$
versus temperature $T$.
Symbols: treatment of observations on iPP (unfilled circles)
and LDPE (filled circles).
Solid lines: approximation of the experimental data
by Eq. (31).
Curve 1: $V_{0}=4.42\cdot 10^{3}$, $V_{1}=16.41$.
Curve 2: $V_{0}=-1.06\cdot 10^{4}$, $V_{1}=40.22$
\vspace*{2 mm}

{\bf Figure 7:}
The standard deviation of activation energies
$\bar{\Sigma}_{\ast}$ versus temperature $T$.
Symbols: treatment of observations on iPP
(unfilled circles) and LDPE (filled circles).
Solid lines: approximation of the experimental data
by Eq. (32).
Curve 1: $\Sigma_{0}=1.01\cdot 10^{3}$, $\Sigma_{1}=1.79$.
Curve 2: $\Sigma_{0}=3.36\cdot 10^{3}$, $\Sigma_{1}=-9.52$
\vspace*{2 mm}

{\bf Figure 8:}
The average relaxation time $\tau_{0}$ versus temperature $T$.
Symbols: treatment of observations on iPP (unfilled circles)
and LDPE (filled circles).
Solid lines: approximation of the experimental data
by Eq. (37).
Curve 1: $\tau_{0}=-13.74$, $\tau_{1}=5.24\cdot 10^{3}$.
Curve 2: $\tau_{0}=-22.18$, $\tau_{1}=9.12\cdot 10^{3}$

\setlength{\unitlength}{0.75 mm}

\begin{figure}[tbh]
\begin{center}

\end{center}
\vspace*{10 mm}

\caption{}
\end{figure}

\begin{figure}[tbh]
\begin{center}
%
\end{center}
\vspace*{10 mm}

\caption{}
\end{figure}

\begin{figure}[tbh]
\begin{center}
%
\end{center}
\vspace*{10 mm}

\caption{}
\end{figure}

\begin{figure}[tbh]
\begin{center}
%
\end{center}
\vspace*{10 mm}

\caption{}
\end{figure}

\begin{figure}[tbh]
\begin{center}
%
\end{center}
\vspace*{10 mm}

\caption{}
\end{figure}

\begin{figure}[tbh]
\begin{center}

\end{center}
\vspace*{10 mm}

\caption{}
\end{figure}

\begin{figure}[tbh]
\begin{center}

\end{center}
\vspace*{10 mm}

\caption{}
\end{figure}

\begin{figure}[tbh]
\begin{center}

\end{center}
\vspace*{10 mm}

\caption{}
\end{figure}

\end{document}